\documentclass[11pt,a4paper]{article}
\pdfoutput=1
\usepackage{jheppub}
\usepackage{tikz}
\usetikzlibrary{intersections, calc, arrows.meta}
\usepackage{subcaption}
\usepackage{bm}

\DeclareMathOperator{\Tr}{Tr}

\newcommand{\ri}{\mathrm{i}}

\newcommand{\hf}{\frac{1}{2}}
\newcommand{\qu}{\frac{1}{4}}

\newcommand{\del}{\partial}

\newcommand{\bra}{\langle}
\newcommand{\ket}{\rangle}
\newcommand{\la}{\lambda}

\newcommand{\h}[1]{\widehat{#1}}

\newcommand{\Ga}{\Gamma}
\newcommand{\al}{\alpha}

\newcommand{\rt}[1]{\sqrt{#1}}
\newcommand{\cO}{\mathcal{O}}

\newcommand{\cN}{\mathcal{N}}
\newcommand{\cM}{\mathcal{M}}

\newcommand{\mZ}{\mathbb{Z}}

\begin{document}

\title{'t Hooft expansion of $SO(N)$ and $Sp(N)$ $\cN=4$ SYM revisited}

\author{Kazumi Okuyama}

\affiliation{Department of Physics, 
Shinshu University, 3-1-1 Asahi, Matsumoto 390-8621, Japan}

\emailAdd{kazumi@azusa.shinshu-u.ac.jp}

\abstract{We study the 't Hooft expansion of 
$d=4$ $\mathcal{N}=4$ supersymmetric Yang-Mills (SYM) theory
with the gauge group
$SO(N)$ or $Sp(N)$.
We consider the $1/N_5$ expansion with 
fixed $g_sN_5$, where 
$g_s$ denotes the string coupling
of bulk type IIB string theory on $AdS_5\times\mathbb{RP}^5$
and $N_5$ refers to the RR 5-form flux through $\mathbb{RP}^5$. $N_5$ 
differs from $N$ due to a shift coming from the RR charge of O3-plane.
As an example, we consider
the $1/N_5$ expansion of the free energy of $\mathcal{N}=4$
SYM on $S^4$ and the $1/2$ BPS circular Wilson loops in the fundamental
representation of $SO(N)$ or $Sp(N)$.
We find that the $1/N_5$ expansion is more ``closed string like''
than the ordinary $1/N$ expansion.}

\maketitle

\section{Introduction}
$d=4$ $\cN=4$ supersymmetric Yang-Mills (SYM) theory
with the gauge group $SO(N)$ or $Sp(N)$ is realized as a worldvolume
theory on D3-branes in the presence of orientifold 3-plane,
and it is holographically dual to the
type IIB string theory on $AdS_5\times\mathbb{RP}^5$ \cite{Witten:1998xy}.
In \cite{Fiol:2014fla,Giombi:2020kvo}, 
the $1/2$ BPS circular Wilson loops
in $SO(N)$ and $Sp(N)$ $\cN=4$ SYM are studied in the
$1/N$ expansion with fixed $g_sN$, where $g_s$ denotes the string coupling
of bulk type IIB string theory. Note that $g_s$ 
is proportional to
the square of the Yang-Mills coupling constant $g_{\text{YM}}^2$.
In \cite{Alday:2021vfb,Dorigoni:2022zcr}, it is suggested that
in the context of AdS/CFT correspondence, it is more natural to
consider the $1/N_5$ expansion with fixed $g_sN_5$, where
$N_5$ refers to the RR 5-form flux through $\mathbb{RP}^5$
in the bulk type IIB string theory
on $AdS_5\times \mathbb{RP}^5$
\begin{equation}
\begin{aligned}
 N_5=\int_{\mathbb{RP}^5}\frac{G_5}{2\pi}.
\end{aligned} 
\end{equation}
As shown in \cite{Blau:1999vz,Aharony:2000cw,Bergman:2001rp},
$N_5$ is shifted from $N$ due to the RR charge of orientifold 
3-plane 
\begin{equation}
N_5=\left\{
\begin{aligned}
 & \frac{N}{2}-\qu,\quad &\text{for}&~~SO(N),\\
&\frac{N}2+\qu ,\quad &\text{for}&~~Sp(N).
\end{aligned} \right.
\label{eq:N5-def}
\end{equation}
Rather surprisingly, the $1/N_5$ expansion with fixed $g_sN_5$ has 
not been fully explored
in the literature before, as far as we know.
In this paper, we will study the $1/N_5$ 
expansion for the partition function of $\cN=4$
SYM on $S^4$ as well as the $1/2$ BPS circular Wilson loops in the fundamental 
representation of $SO(N)$ or $Sp(N)$.\footnote{
Note that $SO(2n)$ gauge theories and
$Sp(2n)$ gauge theories are formally related by the
replacement $n\to-n$ \cite{Mkrtchian:1981bb,Cvitanovic:1982bq}.
}
In the original $1/N$ expansion, both even and odd powers of 
$1/N$ appear in the expansion of the 1/2 BPS Wilson loops
\cite{Fiol:2014fla,Giombi:2020kvo}.
It turns out that the $1/N_5$ expansion is more ``closed string like'':
In the $1/N_5$ expansion of the partition function,
only the even powers of $1/N_5$ appear and in the 
$1/N_5$ expansion of 1/2 BPS Wilson loops only 
the odd powers of $1/N_5$ appear, except for a constant term.
This is consistent with the general property of holography where
D-branes/O-planes are replaced by a closed string background
in the bulk gravitational picture.\footnote{
The question of open versus closed string expansions for the expectation values of Wilson loops of $\cN = 4$ SYM was addressed for $G=SU(N)$ in
\cite{Fiol:2018yuc}.
}

This paper is organized as follows.
In section \ref{sec:vol}, we study the $1/N_5$
expansion of the partition function
of $\cN=4$ SYM, which is inversely
proportional to
the volume of the gauge group $SO(N)$ or $Sp(N)$.
We find that the volume of the gauge group
is characterized by a universal function $V(N_5)$
for both $SO(N)$ and $Sp(N)$, up to an overall factor
$2^{\pm N_5}$.
It turns out that the $1/N_5$ expansion of $\log V(N_5)$ 
contains only even powers of $1/N_5$. 
In section \ref{sec:wilson}, we study the $1/N_5$ expansion of the 
$1/2$ BPS circular Wilson loops of $\cN=4$ SYM in the fundamental representation of
$SO(N)$ or $Sp(N)$.
Finally, we conclude in section \ref{sec:discussion}
with some discussions.
In appendix \ref{app:proof}, we present a proof of the relation \eqref{eq:delg-W}.

\section{$1/N_5$ expansion of the volume of $SO(N)$ and $Sp(N)$}\label{sec:vol}
In this section, we consider the $1/N_5$ expansion of 
the free energy of $\cN=4$ SYM on $S^4$
with the gauge group $G=SO(N)$ or $G=Sp(N)$.
As shown by Pestun \cite{Pestun:2007rz},
the partition function of $\cN=4$ SYM on $S^4$
reduces to a Gaussian matrix model
owing to the supersymmetric localization
\begin{equation}
\begin{aligned}
 Z_G=\frac{1}{\text{vol}(G)}\int_{\text{Lie}(G)}dM e^{-\frac{1}{2g_s}\Tr M^2},
\end{aligned} 
\label{eq:Gaussian}
\end{equation}
where the integral of $M$ is over the Lie algebra of gauge group $G$.
Since the integral of $M$ is Gaussian, the $g_s$-dependence of
$Z_G$ is rather simple
\begin{equation}
\begin{aligned}
 Z_G=\frac{(2\pi g_s)^{\hf\dim G}}{\text{vol}(G)},
\end{aligned} 
\end{equation}
and $Z_G$ is essentially determined by the volume of the gauge group $G$.
Thus, in what follows we will consider the $1/N_5$ expansion of
$\text{vol}(G)$.

The volume of $SO(N)$ is given by \cite{Zyczkowski,Macdonald,Ooguri:2002gx}
\begin{equation}
\begin{aligned}
 \text{vol}\bigl[SO(N)\bigr]&=
\frac{2^{N-\hf}\pi^{\qu N(N+1)}}{\prod_{k=1}^N\Ga(k/2)}
=\left\{
\begin{aligned}
&\frac{2^\hf (2\pi)^{n^2}}{(n-1)!\prod_{i=1}^{n-1}(2i-1)!}, \quad & (N=&2n),\\
&\frac{2^{n+\hf} (2\pi)^{n^2+n}}{\prod_{i=1}^{n}(2i-1)!},\quad & (N=&2n+1).
\end{aligned}\right.
\end{aligned} 
\end{equation}
Our definition of the volume of $SO(2n)$ is the same as
that in \cite{Ooguri:2002gx}, but the volume of
$SO(2n+1)$ differs
from \cite{Ooguri:2002gx} by a factor of $(\pi/2)^\qu$.
The volume of $Sp(N)$ is given by \cite{Ooguri:2002gx}
\begin{equation}
\begin{aligned}
\text{vol}\bigl[Sp(2n)\bigr]&=\frac{2^{-n}(2\pi)^{n^2+n}}{\prod_{i=1}^n(2i-1)!}. 
\end{aligned} 
\end{equation}
From the definition of $N_5$ in \eqref{eq:N5-def},
one can rewrite the above volumes in terms of $N_5$ as
\begin{equation}
\begin{aligned}
 \text{vol}(G)&=2^{\pm N_5}V(N_5),\\
V(N_5)&=
\frac{2^{N_5}\pi^{(N_5+1/4)(N_5+3/4)}G_2(1/2)}{G_2\left(N_5+\frac{3}{4}\right)
G_2\left(N_5+\frac{5}{4}\right)},
\end{aligned}
\label{eq:volG} 
\end{equation}
where the $\pm$ sign corresponds to $G=SO(N)$ and $G=Sp(N)$, respectively,
and $G_2(z)$ denotes the Barnes $G$-function.

Now, let us consider the free energy coming
from the volume of the gauge group $G$
\begin{equation}
\begin{aligned}
 -\log\bigl[\text{vol}(G)\bigr]&=\mp N_5\log2-\log V(N_5).
\end{aligned} 
\label{eq:log2}
\end{equation}
The $1/N_5$ expansion of the Barnes $G$-function in \eqref{eq:volG} 
can be computed by integrating the asymptotic expansion
of the $\Ga$-function\footnote{
See e.g. \href{http://dlmf.nist.gov/5.11.E8}{http://dlmf.nist.gov/5.11.E8}.
}
\begin{equation}
\begin{aligned}
 \log\Ga(z+a)=(z+a-1/2)\log z-z+\hf\log(2\pi)
+\sum_{k=2}^{\infty}\frac{(-1)^kB_k(a)}{k(k-1)z^{k-1}},
\end{aligned} 
\end{equation}
where $B_k(a)$ denotes the Bernoulli polynomial.
After some algebra, we find
\begin{equation}
\begin{aligned}
-\log V(N_5)&=c_0+N_5^2
\left(-\frac{3}{2}+\log\frac{N_5}{\pi}\right)-\frac{5}{48}\log N_5
\\
&+\sum_{g=2}^\infty N_5^{2-2g}\left[\frac{B_{2g}(1/4)}{2g(g-1)}-\frac{B_{2g-1}(1/4)}{4(g-1)(2g-1)}\right],
\end{aligned} 
\label{eq:VN5-exp}
\end{equation}
where $c_0$ is some constant.
As we mentioned in the Introduction,
the $1/N_5$ expansion of $\log V(N_5)$ has only even powers of
$1/N_5$.

This $1/N_5$ expansion \eqref{eq:VN5-exp}
should be compared with the $1/N$ expansion appearing in the topological string.
In the case of topological string, the natural expansion parameter
is $1/N_{\text{top}}$, where $N_{\text{top}}$ is given by \cite{Ooguri:2002gx}
\begin{equation}
\begin{aligned}
 N_{\text{top}}=N\mp 1.
\end{aligned} 
\label{eq:Ntop}
\end{equation}
Here the upper and lower sign correspond to $G=SO(N)$ and $G=Sp(N)$, respectively.
The $1/N_{\text{top}}$ expansion of the volume of $G$ is computed in 
\cite{Ooguri:2002gx}
\begin{equation}
\begin{aligned}
 -\log \bigl[\text{vol}(G)\bigr]
=\hf\sum_g \left(\chi(\cM_{g})N_{\text{top}}^{2-2g}\pm 
\chi(\cM_{g}^1)N_{\text{top}}^{1-2g}\right),
\end{aligned} 
\label{eq:vol-top}
\end{equation}
where $\chi(\cM_{g})$ and $\chi(\cM_{g}^1)$ denote
the Euler characteristic of the moduli space of 
Riemann surfaces of genus $g$
with zero and one cross-cap, respectively \cite{goulden2001}
\begin{equation}
\begin{aligned}
 \chi(\cM_{g})=\frac{B_{2g}}{2g(2g-2)},\quad
\chi(\cM_{g}^1)=\frac{2^{2g-1}B_{2g}(1/2)}{2g(2g-1)}.
\end{aligned} 
\end{equation}
One can see that the $1/N_{\text{top}}$ expansion of $\text{vol}(G)$
contains both even and odd powers of $1/N_{\text{top}}$,
while the $1/N_5$ expansion of $\text{vol}(G)$
contains only even powers of $1/N_5$
except for the first term of \eqref{eq:log2}.
This difference comes from the different definition of 
$N_5$ in \eqref{eq:N5-def} and $N_{\text{top}}$ in \eqref{eq:Ntop}.
As discussed in \cite{Ooguri:2002gx,Sinha:2000ap}, the shift of $N$ in $N_{\text{top}}$
\eqref{eq:Ntop}
comes from the RR charge of topological O-plane, which differs from 
the RR charge of O3-plane in type IIB string theory.
Thus the $1/N_{\text{top}}$ expansion \eqref{eq:vol-top} cannot be applied to 
our case of $\cN=4$ SYM.
In the holographic duality
between $\cN=4$ SYM with the gauge group $SO(N)$ or $Sp(N)$ and
the type IIB string theory on $AdS_5\times \mathbb{RP}^5$,
we should use the $1/N_5$ expansion \eqref{eq:VN5-exp},
instead of the $1/N_{\text{top}}$ expansion \eqref{eq:vol-top}.

\section{$1/2$ BPS Wilson loops in the fundamental representation of $SO(N)$ and $Sp(N)$}\label{sec:wilson}
In this section, we consider the $1/N_5$ expansion of the 
$1/2$ BPS circular Wilson loops in the 
fundamental representation of $G=SO(N)$ or $G=Sp(N)$.
As shown in \cite{Erickson:2000af, Drukker:2000rr, Pestun:2007rz},
the expectation value of the $1/2$ BPS circular Wilson loop is given by
the Gaussian matrix model
\begin{equation}
\begin{aligned}
 W_G=\Bigl\bra\Tr_F e^{M}\Bigr\ket,
\end{aligned} 
\label{eq:mat-exp}
\end{equation}
where the expectation value is defined by the Gaussian measure
\eqref{eq:Gaussian}.
Note that in our definition of $W_G$ we do not divide it by the dimension 
$N$ of 
the fundamental representation.
The Gaussian integral \eqref{eq:mat-exp}
can be evaluated by the method of orthogonal polynomials
and the result is written in terms of the Laguerre polynomials \cite{Fiol:2014fla}
\begin{equation}
\begin{aligned}
W_{SO(2n)}&=2e^{\hf g_s}\sum_{i=0}^{n-1}L_{2i}(-g_s),\\
W_{SO(2n+1)}&=1+2e^{\hf g_s}\sum_{i=0}^{n-1}L_{2i+1}(-g_s),\\
W_{Sp(2n)}&=2e^{\hf g_s}\sum_{i=0}^{n-1}L_{2i+1}(-g_s).
\end{aligned} 
\label{eq:WG-sum}
\end{equation}
One can check that they are correctly normalized as
\begin{equation}
\begin{aligned}
 W_{SO(N)}\Big|_{g_s=0}=N,\quad W_{Sp(N)}\Big|_{g_s=0}=N.
\end{aligned} 
\label{eq:g0-cond}
\end{equation}
As explained in appendix \ref{app:proof},
the derivative of $W_G$ with respect to $g_s$ has a simple form
\begin{equation}
\begin{aligned}
 \del_{g_s}W_{SO(N)}=e^{\hf g_s}L_{N-2}^{(2)}(-g_s),\quad
\del_{g_s}W_{Sp(N)}=e^{\hf g_s}L_{N-1}^{(2)}(-g_s),
\end{aligned} 
\label{eq:delg-W}
\end{equation}
which are both written in terms of $N_5$ as
\begin{equation}
\begin{aligned}
 \del_{g_s} W_G=e^{\hf g_s}L_{2N_5-\frac{3}{2}}^{(2)}(-g_s).
\end{aligned} 
\end{equation}

\subsection{$1/N_5$ expansion of $W_G$}
In this subsection, we consider the $1/N_5$ expansion
of $W_G$ with fixed 't Hooft parameter $\la$
\begin{equation}
\begin{aligned}
 \la=8g_sN_5.
\end{aligned} 
\end{equation}
To do this, it is useful to
express $W_G$ as a contour integral \cite{Okuyama:2006ir}.
Let us first consider $W_{SO(N)}$ for definiteness.
Using the series expansion of the Laguerre polynomial
\begin{equation}
\begin{aligned}
 L_n^{(\al)}(-g_s)=\sum_{i=0}^n\binom{n+\al}{n-i}\frac{g_s^i}{i!},
\end{aligned} 
\end{equation}
$\del_{g_s}W_{SO(N)}$ in \eqref{eq:delg-W} is written as
\begin{equation}
\begin{aligned}
 \del_{g_s} W_{SO(N)}&=
e^{\hf g_s}\sum_{i=0}^{N-2}\binom{N}{N-2-i}\frac{g_s^i}{i!}\\
&=e^{\hf g_s}\sum_{i=0}^{N-2}\frac{N!g_s^i}{(N-2-i)!(i+2)!i!}\\
&=e^{\hf g_s}\oint\frac{dw}{2\pi\ri}\sum_{i=0}^{N-2}\frac{N!}{(N-2-i)!(i+2)!}w^{i-1}
\frac{g_s^i}{i!w^i}\\
&=e^{\hf g_s}\oint\frac{dw}{2\pi\ri}\frac{(1+w)^N}{w^3}e^{\frac{g_s}{w}},
\end{aligned} 
\end{equation}
where the contour of $w$-integral is a circle surrounding $w=0$ counterclockwise.
By the change of variable $w=e^{2z}-1$ we find
\begin{equation}
\begin{aligned}
 \del_{g_s} W_{SO(N)}&=\oint\frac{dz}{2\pi\ri}\frac{1}{4\sinh^3z}e^{(2N-1)z+
\frac{g_s}{2}\coth z}\\
&=\oint\frac{dz}{2\pi\ri}\frac{1}{4\sinh^3z}e^{4N_5z+
\frac{g_s}{2}\coth z},
\end{aligned}
\label{eq:delg-W-z} 
\end{equation}
where the contour of $z$-integral is around $z=0$.
For the $Sp(N)$ case, one can show that $\del_{g_s}W_{Sp(N)}$ is also
given by the same formula
\eqref{eq:delg-W-z}. Thus we find
\begin{equation}
\begin{aligned}
 \del_{g_s} W_{G}=\oint\frac{dz}{2\pi\ri}\frac{1}{4\sinh^3z}e^{4N_5z+
\frac{g_s}{2}\coth z}.
\end{aligned} 
\end{equation}
Finally, integrating this expression with respect to $g_s$, 
we arrive at
\begin{equation}
\begin{aligned}
 W_G=\pm \hf+\oint\frac{dz}{2\pi\ri}\frac{1}{\sinh z\sinh 2z}e^{4N_5z+
\frac{g_s}{2}\coth z}.
\end{aligned} 
\label{eq:WG-oint}
\end{equation}
Here we have determined the integration constant by the normalization
condition 
\eqref{eq:g0-cond}.

In order to study the 't Hooft expansion of
$W_G$, it is convenient to further rewrite
the second term of \eqref{eq:WG-oint} as
\begin{equation}
\begin{aligned}
 &\oint\frac{dz}{2\pi\ri}\frac{1}{\sinh z\sinh 2z}e^{4N_5z+
\frac{g_s}{2}\coth z}\\
=&\oint\frac{dz}{2\pi\ri}
\left(\frac{1}{2\sinh^2z}-\frac{2\sinh^2\frac{z}{2}}{\sinh z\sinh 2z}\right)
e^{4N_5z+\frac{g_s}{2}\coth z}\\
=&\oint\frac{dz}{2\pi\ri}\left(-\frac{1}{g_s}e^{4N_5z}
\del_z e^{\frac{g_s}{2}\coth z}\right)-
\oint\frac{dz}{2\pi\ri}\frac{2\sinh^2\frac{z}{2}}{\sinh z\sinh 2z}
e^{4N_5z+\frac{g_s}{2}\coth z}\\
=&\frac{4N_5}{g_s}\oint\frac{dz}{2\pi\ri}e^{4N_5z+\frac{g_s}{2}\coth z}
-\oint\frac{dz}{2\pi\ri}\frac{2\sinh^2\frac{z}{2}}{\sinh z\sinh 2z}
e^{4N_5z+\frac{g_s}{2}\coth z}.
\end{aligned} 
\label{eq:w-ointz}
\end{equation}
One can show that the first term
of \eqref{eq:w-ointz} is equal to the $1/2$ BPS Wilson loop
of $U(2N_5)$ $\cN=4$ SYM \cite{Drukker:2000rr}
\begin{equation}
\begin{aligned}
 \frac{4N_5}{g_s}\oint\frac{dz}{2\pi\ri}e^{4N_5z+\frac{g_s}{2}\coth z}=e^{\hf g_s}
L_{2N_5-1}^{(1)}(-g_s)=W_{U(2N_5)}.
\end{aligned} 
\label{eq:WU}
\end{equation}

Let us consider the 't Hooft expansion of $W_{U(2N_5)}$ in \eqref{eq:WU}
following the approach of \cite{Okuyama:2006ir}.
By rescaling $z\to g_sz$, $W_{U(2N_5)}$ is written as
\begin{equation}
\begin{aligned}
 W_{U(2N_5)}&=4N_5 \oint\frac{dz}{2\pi\ri}e^{\hf (\la z+z^{-1})+\frac{g_s}{2}\coth g_sz-\hf z^{-1}}.
\end{aligned} 
\label{eq:W-UN}
\end{equation}
The first part of the exponential 
$e^{\hf (\la z+z^{-1})}$ is essentially the generating function of the 
modified Bessel function of the first kind $I_n(x)$
\begin{equation}
\begin{aligned}
 e^{\hf (\la z+z^{-1})}=\sum_{n\in\mZ}\frac{\h{I}_n}{z^n},\qquad
\end{aligned} 
\label{eq:Bessel-gen}
\end{equation}
where $\h{I}_n$ is given by
\begin{equation}
\begin{aligned}
\h{I}_n=\frac{I_n(\rt{\la})}{(\rt{\la})^n}.
\end{aligned} 
\end{equation}
The second part of the the exponential in \eqref{eq:W-UN}
can be expanded in $g_s$ as
\begin{equation}
\begin{aligned}
 e^{\frac{g_s}{2}\coth g_sz-\hf z^{-1}}=1+\frac{z}{6}g_s^2+
\left(\frac{z^2}{72}-\frac{z^3}{90}\right)g_s^4+
\cO(g_s^6).
\end{aligned} 
\end{equation}
Then, taking the residue at $z=0$ we find the small $g_s$
expansion of $W_{U(2N_5)}$
\begin{equation}
\begin{aligned}
 W_{U(2N_5)}=\frac{\la}{2g_s}
\left[\h{I}_1+\frac{\h{I}_2}{6}g_s^2+
\left(\frac{\h{I}_3}{72}-\frac{\h{I}_4}{90}\right)g_s^4
+\cO(g_s^6)\right].
\end{aligned} 
\label{eq:WU5-exp}
\end{equation}
Note that the small $g_s$ expansion with fixed $\la=8g_sN_5$
is basically the same as the $1/N_5$ expansion since $g_s$ and
$1/N_5$ are related by
\begin{equation}
\begin{aligned}
 g_s=\frac{\la}{8N_5}.
\end{aligned} 
\end{equation}

Next consider the second term of \eqref{eq:w-ointz}, which we will
denote by $W_{T}$
\begin{equation}
\begin{aligned}
 W_{T}&=-\frac{g_s}{4}\oint\frac{dz}{2\pi\ri}\frac{8\sinh^2\frac{g_sz}{2}}{\sinh g_sz\sinh 2g_sz}
e^{\hf(\la z+z^{-1})+\frac{g_s}{2}\coth g_sz-\hf z^{-1}}.
\end{aligned} 
\end{equation}
Again, the first part of the exponential has the expansion
\eqref{eq:Bessel-gen} and the rest of the integrand can be expanded in $g_s$ as
\begin{equation}
\begin{aligned}
 \frac{8\sinh^2\frac{g_sz}{2}}{\sinh g_sz\sinh 2g_sz}
e^{\frac{g_s}{2}\coth g_sz-\hf z^{-1}}&=1+\left(\frac{z}{6}-\frac{3z^2}{4}\right)g_s^2
\\
&\qquad+\left(\frac{z^2}{72}-\frac{49z^3}{360}+\frac{3z^4}{8}\right)g_s^4+\cO(g_s^6).
\end{aligned} 
\end{equation}
Taking the residue at $z=0$, we find
the small $g_s$ expansion of $W_{T}$ with fixed $\la$
\begin{equation}
\begin{aligned}
 W_{T}&=-\frac{g_s}{4}\left[\h{I}_1+
\left(\frac{\h{I}_2}{6}-\frac{3\h{I}_3}{4}\right)g_s^2+
\left(\frac{\h{I}_3}{72}-\frac{49\h{I}_4}{360}+\frac{3\h{I}_5}{8}\right)g_s^4
+\cO(g_s^6)\right].
\end{aligned} 
\label{eq:WO-exp}
\end{equation}

To summarize, we find that $W_G$ is decomposed as
\begin{equation}
\begin{aligned}
 W_G=\pm\hf +W_{U(2N_5)}+W_{T},
\end{aligned} 
\label{eq:WG-decomp}
\end{equation}
and the last two terms are expanded as
\begin{equation}
\begin{aligned}
 W_{U(2N_5)}&=\sum_{g=0}^\infty a_g(\la) g_s^{2g-1}
=\sum_{g=0}^\infty a_g(\la) \left(\frac{\la}{8N_5}\right)^{2g-1},\\
W_{T}&=\sum_{g=0}^\infty b_g(\la) g_s^{2g+1}
=\sum_{g=0}^\infty b_g(\la)\left(\frac{\la}{8N_5}\right)^{2g+1},
\end{aligned} 
\end{equation}
where $a_g(\la)$ and $b_g(\la)$ are some functions of $\la$
whose explicit forms can be found in \eqref{eq:WU5-exp} and \eqref{eq:WO-exp}.
One can see that $W_{U(2N_5)}$ and $W_{T}$ are both expanded in $1/N_5$
with only odd powers of $1/N_5$.

\subsection{Relation to the ordinary $1/N$ 't Hooft expansion}
Let us compare our $1/N_5$ expansion of $W_{G}$ with 
the ordinary $1/N$ expansion of $W_{G}$. 
For definiteness, we consider the $G=SO(N)$ case.
The $1/N$ expansion of 
$W_{SO(N)}$ is studied in \cite{Fiol:2014fla,Giombi:2020kvo} where the 
't Hooft coupling $\la'$ is defined as
\begin{equation}
\begin{aligned}
 \la'=8g_sN.
\end{aligned} 
\end{equation}
To this end, it is convenient to start with the 
expression of $W_{SO(N)}$ found in \cite{Fiol:2014fla} 
\footnote{See also appendix \ref{app:proof} for 
a derivation of this expression.}
\begin{equation}
\begin{aligned}
 W_{SO(N)}&=e^{\hf g_s}L_{N-1}^{(1)}(-g_s)-\hf\int_0^{g_s}dx
\,e^{\hf x}L_{N-1}^{(1)}(-x)\\
&=W_{U(N)}(g_s)-\hf \int_0^{g_s}dx\,W_{U(N)}(x).
\end{aligned} 
\label{eq:W-fiol}
\end{equation}
From the known $1/N$ expansion of the $1/2$ BPS Wilson loop in $U(N)$
$\cN=4$ SYM \cite{Drukker:2000rr},
one can easily compute the $1/N$ expansion of $W_{SO(N)}$
\begin{equation}
\begin{aligned}
 W_{SO(N)}=\hf+\frac{2\rt{2}N}{\rt{\la'}}I_1(\rt{\la'/2})
-\hf I_0(\rt{\la'/2})+\frac{\la'I_2(\rt{\la'/2})}{96N}
-\frac{\la'^{3/2}I_3(\rt{\la'/2})}{384\rt{2}N^2}
+\cO(N^{-3}).
\end{aligned} 
\label{eq:w-giombi}
\end{equation}
One can check that the $1/N$ expansion
in \eqref{eq:w-giombi} and our $1/N_5$ expansion 
are related by the change of parameters
$(\la',N)\to(\la,g_s)$
\begin{equation}
\begin{aligned}
\la'=2\la+4g_s,\quad N=\frac{\la}{4g_s}+\hf.
\end{aligned} 
\end{equation}
Plugging this relation into \eqref{eq:w-giombi}
and expanding in $g_s$, we find
\begin{equation}
\begin{aligned}
 W_{SO(N)}=\hf+\frac{\la}{2g_s}\left[\h{I}_1
+\frac{\h{I}_2}{6}g_s^2\right]
-\frac{g_s}{4}\h{I}_1+\cO(g_s^3).
\end{aligned}
\label{eq:giombi-rewrite} 
\end{equation}
This agrees with our result of $1/N_5$ expansion
\eqref{eq:WU5-exp} and \eqref{eq:WO-exp}
up to this order $\cO(g_s^3)$, as expected.

Note that, 
in the original $1/N$ expansion \eqref{eq:w-giombi} both even and odd powers of
$N^{-1}$ appear. On the other hand, 
in our case \eqref{eq:WG-decomp} only the odd powers of $g_s$
arise, except for the constant term $\pm 1/2$ in \eqref{eq:WG-decomp}.
Although our decomposition \eqref{eq:WG-decomp}
is similar to \eqref{eq:W-fiol}, we stress that they are 
different. In particular, our $W_{T}$ is not equal to the second term of 
\eqref{eq:W-fiol}.

\section{Conclusions and outlook}\label{sec:discussion}
In this paper, we have studied the $1/N_5$ expansion of the 
volume of the gauge group $G$
and the $1/2$ BPS Wilson loops in the fundamental representation of $G$ in
$\cN=4$ SYM with $G=SO(N)$ or $G=Sp(N)$ .
Due to the shift of $N$ coming from the RR charge of O3-plane \eqref{eq:N5-def},
the $1/N_5$ expansion with fixed 't Hooft parameter $\la=8g_sN_5$
is different from the ordinary $1/N$ expansion. We found that the $1/N_5$ expansion
looks more ``closed string like''
than the ordinary $1/N$ expansion.
For instance, we found that the $1/N_5$ expansion
of the volume of $G$ contains only the even powers of $1/N_5$, except
for the first term $\mp N_5\log2$ in \eqref{eq:log2}.
This is different from the $1/N_{\text{top}}$ expansion
of $\text{vol}(G)$ in topological string
\cite{Ooguri:2002gx}.
It would be interesting to find a mathematical meaning, if any, 
of the coefficient
of $N_5^{2-2g}$ in \eqref{eq:VN5-exp}
as a certain quantity on the moduli space of Riemann surfaces of genus $g$.

We have also studied the $1/N_5$ expansion of the $1/2$ BPS Wilson loop
$W_G$
in the fundamental representation of $G=SO(N)$ or $G=Sp(N)$.
We found that $W_G$ is decomposed as \eqref{eq:WG-decomp}.
Except for the constant term $\pm1/2$ in \eqref{eq:WG-decomp}, 
$W_{U(2N_5)}$ and $W_{T}$ are both expanded 
in $1/N_5$ with only odd powers of $1/N_5$.
It is tempting to speculate that $W_{U(2N_5)}$ and $W_{T}$
correspond to the untwisted and the twisted sector of bulk type IIB string theory
on $AdS_5\times\mathbb{RP}^5$.
It would be interesting to understand the bulk gravitational interpretation of the
decomposition \eqref{eq:WG-decomp} more clearly.

It would be interesting to extend our analysis to more general
observables in $\cN=4$ SYM with the gauge group $SO(N)$ or $Sp(N)$,
such as an integrated four-point correlator \cite{Dorigoni:2022zcr} and
the $1/2$ BPS Wilson loop in the spinor representation of $SO(N)$
\cite{Fiol:2014fla,Giombi:2020kvo}, to name a few.
We leave this as an interesting future problem.

\acknowledgments
This work was supported
in part by JSPS Grant-in-Aid for Transformative Research Areas (A) 
``Extreme Universe'' No. 21H05187 and JSPS KAKENHI Grant No. 22K03594.

\appendix
\section{Proof of \eqref{eq:delg-W}}\label{app:proof}
In this appendix, we present a proof of the relation
\eqref{eq:delg-W}. For definiteness we consider $W_{SO(2n)}$.
To this end, we can use the fact that 
the Laguerre polynomial is written as a matrix element of 
the harmonic oscillator (see e.g. \cite{Okuyama:2018yep})
\begin{equation}
\begin{aligned}
 \bra i|e^{\rt{g_s}(a+a^\dag)}|j\ket=
\bra j|e^{\rt{g_s}(a+a^\dag)}|i\ket=\rt{\frac{i!}{j!}}g_s^{\frac{j-i}{2}}L_{i}^{(j-i)}(-g_s),
\end{aligned} 
\label{eq:mat-ele}
\end{equation}
where
\begin{equation}
\begin{aligned}
  {[}a,a^\dag{]}=1,\quad a|0\ket=0,\quad |k\ket=\frac{(a^\dag)^k}{\rt{k!}}|0\ket.
\end{aligned} 
\end{equation}
Then $W_{SO(2n)}$ in \eqref{eq:WG-sum} is written as
\begin{equation}
\begin{aligned}
 W_{SO(2n)}&=2\sum_{i=0}^{n-1}\bra 2i|e^{\rt{g_s}(a+a^\dag)}|2i\ket\\
&=\sum_{k=0}^{2n-1}\bigl[1+(-1)^k\bigr]
\bra k|e^{\rt{g_s}(a+a^\dag)}|k\ket.
\end{aligned} 
\end{equation}
Thus $W_{SO(2n)}$ is naturally decomposed as
\begin{equation}
\begin{aligned}
 W_{SO(2n)}=W_{SO(2n)}^{+}+W_{SO(2n)}^{-},
\end{aligned} 
\end{equation}
where
\begin{equation}
\begin{aligned}
 W_{SO(2n)}^{+}&=\sum_{k=0}^{2n-1}\bra k|e^{\rt{g_s}(a+a^\dag)}|k\ket,\\
W_{SO(2n)}^{-}&=\sum_{k=0}^{2n-1}(-1)^k
\bra k|e^{\rt{g_s}(a+a^\dag)}|k\ket.
\end{aligned} 
\end{equation}
Note that $W_{SO(2n)}^{+}$ is equal to the Wilson loop of 
$U(2n)$ $\cN=4$ SYM \cite{Drukker:2000rr}.
The sum over $k$ in $W_{SO(2n)}^{+}$
can be simplified as
\begin{equation}
\begin{aligned}
 \rt{g_s}W_{SO(2n)}^{+}&=\sum_{k=0}^{2n-1}\bra k|\bigl[a,e^{\rt{g_s}(a+a^\dag)}\bigl]|k\ket\\
&=\sum_{k=0}^{2n-1}\Bigl[\rt{k+1}\bra k+1|e^{\rt{g_s}(a+a^\dag)}|k\ket-
\rt{k}\bra k|e^{\rt{g_s}(a+a^\dag)}|k-1\ket\Bigr]\\
&=\rt{2n}\bra 2n|e^{\rt{g_s}(a+a^\dag)}|2n-1\ket\\
&=\rt{g_s}e^{\hf g_s}L_{2n-1}^{(1)}(-g_s).
\end{aligned} 
\end{equation}
In the last step we used \eqref{eq:mat-ele}.
Thus we find
\begin{equation}
\begin{aligned}
W_{SO(2n)}^{+}=W_{U(2n)} =e^{\hf g_s}L_{2n-1}^{(1)}(-g_s),
\end{aligned} 
\end{equation}
which agrees with the known result of $W_{U(2n)}$ in 
\cite{Drukker:2000rr}. 

Next, let us consider the $g_s$-derivative of $W_{SO(2n)}^{-}$
\begin{equation}
\begin{aligned}
 \del_{g_s}W_{SO(2n)}^{-}&=\frac{1}{2\rt{g_s}}
\sum_{k=0}^{2n-1}(-1)^k\bra k|e^{\rt{g_s}(a+a^\dag)}(a+a^\dag)|k\ket\\
&=\frac{1}{2\rt{g}}
\sum_{k=0}^{2n-1}(-1)^k\Bigl[\rt{k}\bra k|e^{\rt{g_s}(a+a^\dag)}|k-1\ket+
\rt{k+1}\bra k|e^{\rt{g_s}(a+a^\dag)}|k+1\ket\Bigr]\\
&=\frac{1}{2\rt{g_s}}
\sum_{k=0}^{2n-1}\Bigl[(-1)^k\rt{k}\bra k|e^{\rt{g_s}(a+a^\dag)}|k-1\ket
-(-1)^{k+1}\rt{k+1}\bra k+1|e^{\rt{g_s}(a+a^\dag)}|k\ket\Bigr]\\
&=-\frac{1}{2\rt{g_s}}\rt{2n}\bra 2n|e^{\rt{g_s}(a+a^\dag)}|2n-1\ket\\
&=-\hf e^{\hf g_s}L_{2n-1}^{(1)}(-g_s).
\end{aligned} 
\end{equation}
Finally, we find
\begin{equation}
\begin{aligned}
 \del_{g_s}W_{SO(2n)}&=\del_{g_s}W_{SO(2n)}^{+}+\del_{g_s}W_{SO(2n)}^{-}\\
&=\del_{g_s}\Bigl[e^{\hf g_s}L_{2n-1}^{(1)}(-g_s)\Bigr]-\hf e^{\hf g_s}L_{2n-1}^{(1)}(-g_s)\\
&=e^{\hf g_s}L_{2n-2}^{(2)}(-g_s).
\end{aligned} 
\end{equation}
This proves \eqref{eq:delg-W} for the $SO(2n)$ case.
$SO(2n+1)$ and $Sp(N)$ cases can be proved in a similar manner.
\bibliography{paper}
\bibliographystyle{utphys}

\end{document}